\documentclass[prd,twocolumn,showpacs,preprintnumbers,amsmath,amssymb]{revtex4}
\usepackage{graphicx}

\usepackage{bm}

\newcommand{\beq}{\begin{equation}}
\newcommand{\eeq}{\end{equation}}
\newcommand{\beqs}{\begin{eqnarray}}
\newcommand{\eeqs}{\end{eqnarray}}
\newcommand{\lsim}{\mathrel{\raisebox{-
.6ex}{$\stackrel{\textstyle<}{\sim}$}}}

\begin{document}

\title{Standard-Model Condensates and the Cosmological Constant} 

\author{Stanley J. Brodsky$^{a,b}$}

\author{Robert Shrock$^b$}

\affiliation{(a) 
Stanford Linear Accelerator Center, Stanford University, Stanford, CA  94309}

\affiliation{(b)  
C.N. Yang Institute for Theoretical Physics, Stony Brook University, 
Stony Brook, NY 11794}

\begin{abstract}

We suggest a solution to the problem of some apparently excessive contributions
to the cosmological constant from Standard-Model condensates.

\end{abstract}


\maketitle

One of the most challenging problems in physics is that of the cosmological
constant $\Lambda$ (recent reviews include \cite{peeblesratra}-\cite{detf}.
This enters in the Einstein gravitational field equations as \cite{bks,units}
\beq
R_{\mu\nu}-\frac{1}{2}g_{\mu\nu}R-\Lambda g_{\mu\nu} = (8\pi G_N)T_{\mu\nu},
\label{einsteineq}
\eeq
where $R_{\mu\nu}$, $R$, $g_{\mu\nu}$, $T_{\mu\nu}$, and $G_N$ are
the Ricci curvature tensor, the scalar curvature, the metric
tensor, the stress-energy tensor, and Newton's constant.  
One defines 
\beq
\rho_\Lambda = \frac{\Lambda}{8 \pi G_N}
\label{rholambda}
\eeq
and 
\beq
\Omega_\Lambda = \frac{\Lambda}{3H_0^2} = \frac{\rho_\Lambda}{\rho_c}~,
\label{omegalambda}
\eeq
where 
\beq
\rho_c = \frac{3H_0^2}{8\pi G_N} \ , 
\label{rhoc}
\eeq
and $H_0=(\dot a/a)_0$ is the Hubble constant in the present era, with $a(t)$
being the Friedmann-Robertson-Walker scale parameter \cite{bks,cosm}.  Long
before the current period of precision cosmology, it was known that
$\Omega_\Lambda$ could not be larger than O(1). In the context of quantum field
theory, this was very difficult to understand, because estimates of the
contributions to $\rho_\Lambda$ from (i) vacuum condensates of quark and gluon
fields in quantum chromodynamics (QCD) and the vacuum expectation value 
of the Higgs field hypothesized in the Standard Model (SM) to be responsible
for electroweak symmetry breaking, and from (ii) zero-point energies of
quantum fields appear to be too large by many orders of magnitude.
Observations of supernovae showed the accelerated expansion of the universe and
are consistent with the hypothesis that this is due to a cosmological constant,
$\Omega_\Lambda \simeq 0.76$ \cite{hzst,scp,cosmparam}.

   Here we shall propose a solution to the problem of QCD condensate
contributions to $\rho_\Lambda$.  We also comment on other contributions of
type (i) and (ii). Two important condensates in QCD are the quark condensates
$\langle \bar q q \rangle \equiv \langle \sum_{a=1}^{N_c} \bar q_a q^a
\rangle$, where $q$ is a quark whose current-quark mass is small compared with
the confinement scale $\Lambda_{QCD} \simeq 250$ MeV, and the gluon condensate,
$\langle G_{\mu\nu}G^{\mu\nu} \rangle \ \equiv \ \langle \sum_{a=1}^{N_c^2-1}
G^a_{\mu \nu} G^{a \, \mu\nu}\rangle$, where $G^a_{\mu\nu} = \partial_\mu
A^a_\nu - \partial_\nu A^a_\mu + g_s c_{abc} A_\mu^b A_\nu^c$, $a,b,c$ denote
the color indices, $g_s$ is the color SU(3)$_c$ gauge coupling, $N_c=3$, and
$c_{abc}$ are the structure constants for SU(3)$_c$.  These condensates form at
times of order $10^{-5}$ sec. in the early universe as the temperature $T$
decreases below the confinement-deconfinent temperature $T_{dec} \simeq 200$
MeV.  For $T << T_{dec}$, in the conventional quantum field theory view, these
condensates are considered to be constants throughout space.  If this were
true, then they would contribute $(\delta \rho_\Lambda)_{QCD} \sim
\Lambda_{QCD}^4$, so that $(\delta \Omega_\Lambda)_{QCD} \simeq 10^{45}$.
However, we have argued in Ref. \cite{cond} that, contrary to this conventional
view, these condensates (and also higher-order ones such as $\langle (\bar q
q)^2\rangle$ and $\langle (\bar q q) G_{\mu\nu}G^{\mu\nu}\rangle$) have spatial
support within hadrons, not extending throughout all of space.  The reason for
this is that the condensates arise because of quark and gluon interactions, and
these particles are confined within hadrons \cite{cm}.  We have argued that,
consequently, these QCD condensates should really be considered as comprising
part of the masses of hadrons.  Hence, we conclude that their effect on gravity
is already included in the baryon term $\Omega_b$ in $\Omega_m$ and, as such,
they do not contribute to $\Omega_\Lambda$.

Another excessive type-(i) contribution to $\rho_\Lambda$ is conventionally
viewed as arising from the vacuum expectation value of the Standard-Model Higgs
field, $v_{EW} = 2^{-1/4}G_F^{-1/2} = 246$ GeV, giving $(\delta
\rho_\Lambda)_{EW} \sim v_{EW}^4$ and hence $(\delta \Omega_\Lambda)_{EW} \sim
10^{56}$.  Similar numbers are obtained from Higgs vacuum expectation values in
supersymmetric extensions of the Standard Model (recalling that the
supersymmetry breaking scale is expected to be the TeV scale).  However, it is
possible that electroweak symmetry breaking is dynamical; for example, it may
result from the formation of a bilinear condensate of fermions $F$ (called
technifermions) subject to an asymptotically free, vectorial, confining gauge
interaction, commonly called technicolor (TC), that gets strong on the TeV
scale \cite{tc}. In such theories there is no fundamental Higgs field.
Technicolor theories are challenged by, but may be able to survive, constraints
from precision electroweak data.  By our arguments in \cite{cond}, in a
technicolor theory, the technifermion and technigluon condensates would have
spatial support in the technihadrons and techniglueballs and would contribute
to the masses of these states.  We stress that, just as was true for the QCD
condensates, these technifermion and technigluon condensates would not
contribute to $\rho_\Lambda$.  Hence, if a technicolor-type mechanism should
turn out to be responsible for electroweak symmetry breaking, then there would
not be any problem with a supposedly excessive contribution to $\rho_\Lambda$
for a Higgs vacuum expectation value.  Indeed, stable technihadrons in certain
technicolor theories may be viable dark-matter candidates \cite{tcdm}.

We next comment briefly on type-(ii) contributions. The formal expression for
the energy density $E/V$ due to zero-point energies of a quantum field
corresponding to a particle of mass $m$ is 
\beq
E/V = \int \frac{d^3k}{(2\pi)^3} \frac{\omega(k)}{2} \ , 
\label{zeropoint} 
\eeq
where the energy is $\omega(k) = \sqrt{{\bf k}^2 + m^2}$.  However, first, this
expression is unsatisfactory, because it is (quartically) divergent.  In modern
particle physics one would tend to regard this divergence as indicating that
one is using an low-energy effective field theory, and one would impose an
ultraviolet cutoff $M_{UV}$ on the momentum integration, reflecting the upper
range of validity of this low-energy theory.  Since neither the left- nor
right-hand side of eq. (\ref{zeropoint}) is Lorentz-invariant, this cutoff
procedure is more dubious than the analogous procedure for Feynman integrals of
the form $\int d^4k \, I(k,p)$ in quantum field theory, where
$I(k,p_1,...,p_n)$ is a Lorentz-invariant integrand function depending on some
set of 4-momenta $p_1,...,p_n$.  If, nevertheless, one proceeds to use such a
cutoff, then, since a mass scale characterizing quantum gravity (QG) is
$M_{Pl}=G_N^{-1/2}=1.2 \times 10^{19}$ GeV, one would infer that $(\delta
\rho_\Lambda)_{QG} \sim M_{Pl}^4/(16\pi^2)$, and hence $(\delta
\Omega_\Lambda)_{QG} \sim 10^{120}$.  With the various mass scales
characterizing the electroweak symmetry breaking and particle masses in the
Standard Model, one similarly would obtain $(\delta \Omega_\Lambda)_{SM} \sim
10^{56}$.  Given the fact that eq. (\ref{zeropoint}) is not Lorentz-invariant,
one may well question the logic of considering it as a contribution to the
Lorentz-invariant quantity $\rho_\Lambda$ \cite{prcrit,jaffe05}. Indeed, one
could plausibly argue that, as an energy density, it should instead be part of
$T_{00}$ in the energy-momentum tensor $T_{\mu\nu}$.  Phrased in a different
way, if one argues that it should be associated with the $\Lambda g_{\mu \nu}$
term, then there must be a negative corresponding zero-point pressure
satisfying $p=-\rho$, but the source for such a negative pressure is not
evident in eq. (\ref{zeropoint}).

The light-front (LF) quantization of the Standard Model provides another
perspective. In this case, the Higgs field has the form~\cite{lf} $\phi= \omega
+ \varphi$ where $\omega$ is a classical zero mode determined by minimizing the
Yukawa potential $V(\phi)$ of the SM Lagrangian, and $\varphi$ is the quantized
field which creates the physical Higgs particle.  The coupling of the leptons,
quarks, and vector bosons to the zero mode $\omega$ give these particles their
masses. The electroweak phenomenology of the LF-quantized Standard Model is in
fact identical to the usual formulation~\cite{lf}.  In contrast to conventional
instant-form Standard Model is trivial in the light-front
formulation~\cite{lfrev,scottish}, and there is no zero-point fluctuation in
the light-front theory since $\omega$ is a classical quantity. Although this
eliminates any would-be type-(ii) contributions of zero-point fluctuations to
the cosmological constant, the contribution to the electroweak action from the
Standard Model Yukawa potential $V(\omega)$ evaluated at its minimum would, as
in the conventional analysis, yield an excessively large type-(i) contribution
$(\delta \Omega_\Lambda)_{EW} \sim 10^{56}$.  Thus the light-front formulation
of the Standard Model based on a fundamental elementary Higgs field evidently
does not solve the problem with type-(i) electroweak contributions to
$\Omega_\Lambda$. However, as we have noted above, theories with dynamical
electroweak symmetry breaking, such as technicolor, are able to solve the
problem with type-(i) contributions.

In summary, we have suggested a solution to what has hitherto commonly been
regarded as an excessively large contribution to the cosmological constant by
QCD condensates. We have argued that these condensates do not, in fact,
contribute to $\Omega_\Lambda$; instead, they have spatial support within
hadrons and, as such, should really be considered as contributing to the masses
of these hadrons and hence to $\Omega_b$. We have also suggested a possible
solution to what would be an excessive contribution to $\Omega_\Lambda$ from a
hypothetical Higgs vacuum expectation value; the solution would be applicable
if electroweak symmetry breaking occurs via a technicolor-type mechanism.

This research was partially supported by grants DE-AC02-76SF00515 (SJB) and
NSF-PHY-06-53342 (RS).  Preprint SLAC-PUB-13166, YITP-SB-08-09.


\begin{thebibliography}{99}
\bibitem{units}
We use units where $\hbar=c=1$, and our flat-space metric is 
$\eta_{\mu\nu}={\rm diag}(1,-1,-1,-1)$.
\bibitem{cosm} 
The field equations imply $(\dot a/a)^2 = H^2 = (8\pi G_N/3)\rho
+ \Lambda/3 - k/a^2$ and $\ddot a/a = -4\pi G_N (\rho+3p)+\Lambda/3$, where
$\rho=$ total mass/energy density, $p=$ pressure, and $k$ is the
curvature parameter; equivalently, $1 = \Omega_m + \Omega_\gamma +
\Omega_\Lambda + \Omega_k$, where $\Omega_m = 8\pi G_N \rho_m/H_0^2$, 
 $\Omega_\gamma = 8\pi G_N \rho_\gamma/H_0^2$, and 
$\Omega_k = -k/(H_0^2 a^2)$.
\bibitem{cosmparam} 
%
The supernovae data \cite{hzst,scp}, together with measurements of the cosmic
microwave background radiation, galaxy clusters, and other inputs, e.g.,
primordial element abundances, have led to a consistent determination of the
cosmological parameters \cite{wmap}-\cite{pdg}.  These include $H_0 = 73 \pm 3$
km/s/Mpc, $\rho_c = 0.56 \times 10^{-5}$ GeV/cm$^3$ $= (2.6 \times 10^{-3} \
{\rm eV} \,)^4$, total $\Omega_m \simeq 0.24$ with baryon term $\Omega_b \simeq
0.042$, so that the dark matter term is $\Omega_{dm} \simeq 0.20$.  In the
equation of state $p=w\rho$ for the ``dark energy'', $w$ is consistent with
being equal to $-1$, the value if the accelerated expansion is due to a
cosmological constant.  Other suggestions for the source of the accelerated
expansion include modifications of general relativity and time-dependent
$w(t)$, as reviewed in \cite{peeblesratra}-\cite{detf}.
\bibitem{cm}
%
As we noted \cite{cond}, this is somewhat analogous to the fact that although
formal statistical mechanics treatments of condensates and related order
parameters in condensed matter physics take an idealized thermodynamic
(infinite-volume) limit and obtain condensates that are constant throughout
this infinite spatial volume, actual condensates and order parameters are
experimentally measured to exist in, not outside of, the finite-volume samples
wherein the particles responsible for them reside, such as a magnetization 
in a piece of iron.
\bibitem{qm}
%
In accord with the Heisenberg uncertainty principle, a TC condensate also 
affects virtual processes occurring over times $t \lsim 1/300$ \, 
GeV$^{-1}$.  This is relevant for the generation of quark and lepton masses 
in TC/ETC theories.
\bibitem{prcrit}
This criticism of the conventional lore has also been made in 
\cite{peeblesratra} and \cite{jaffe05}.
\end{thebibliography}
\end{document}